\documentclass[letterpaper, 12pt]{article}
\usepackage{graphicx} 
\usepackage{geometry}
\usepackage{enumerate}
\usepackage{verbatim}
\usepackage{fullpage}
\usepackage{amssymb}
\usepackage{amsmath}
\usepackage{pdfpages}
\usepackage{multicol}
\pdfoutput=1

\usepackage{natbib}
\usepackage{indentfirst}
\usepackage{setspace}
\usepackage{float}
\usepackage{multirow}

\title{Bayesian Projection of Life Expectancy Accounting for the HIV/AIDS
Epidemic}
\author{Jessica Godwin and Adrian E. Raftery\\
University of Washington}
\date{August 25, 2016}

\begin{document}
\maketitle

\section{Introduction}
\label{sec:Intro}

Probabilistic projections of mortality measures are important for many applications including population projection and pension and healthcare planning. 
Until recently, most projections of mortality measures were deterministic, although the UN has recently started to base its official projections of mortality on probabilistic methods. Most projections, deterministic or probabilistic, do not incorporate cause-of-death information or other covariates.\par
 \citet{leecarter:1992} developed the first method for projecting a mortality measure probabilistically. The Lee-Carter method projects age-specific mortality rates and is widely used today. It requires at least three time periods of age-specific death rates, an amount of data that is not available in many countries. 
The method assumes that the logarithm of the age-specific death rates will increase linearly in the future, which may not be optimal for long term projections \citep{lee:2001:evaluatingLeeCarter}. \citet{girosi:2008:demforecasting} proposed a Bayesian method for smoothing age-specific death rates over both age and time. Though this method allows for the incorporation of covariates, it has been shown to perform well only for countries with good vital registration data. Like the Lee-Carter method, the \citet{girosi:2008:demforecasting} method assumes a constant rate of increase.
\citet{lutz:1998:expert} developed an expert-based method for probabilistic projections of population that incorporates subjective probabilistic projections for several demographic measures, including life expectancy. 
 
 \par \citet{raftery:2013:e0, raftery:2014:jointe0} presented a Bayesian hierarchical model (BHM) for projecting male and female life expectancy probabilistically for all countries of the world to 2100, and this method is now used by the UN as an input to its official population projections \citep{un:WPP2015}.
We extend the model in \citet{raftery:2013:e0} to include covariate information about generalized HIV epidemic prevalence and coverage of antiretroviral therapy (ART) in each country. A country is said to have a generalized HIV/AIDS epidemic when HIV prevalence is greater than 1\% in the general population, and it is not concentrated in at risk subgroups. While there are many diseases that have a high impact on mortality in a given country, the generalized HIV epidemic is unusual in that it dramatically increases age-specific mortality rates at prime adult ages. Its demographic impact is therefore different from that of other diseases, which tend primarily to affect mortality rates for very young and/or older ages. 

\begin{figure}
\centering
\includegraphics[scale=0.45]{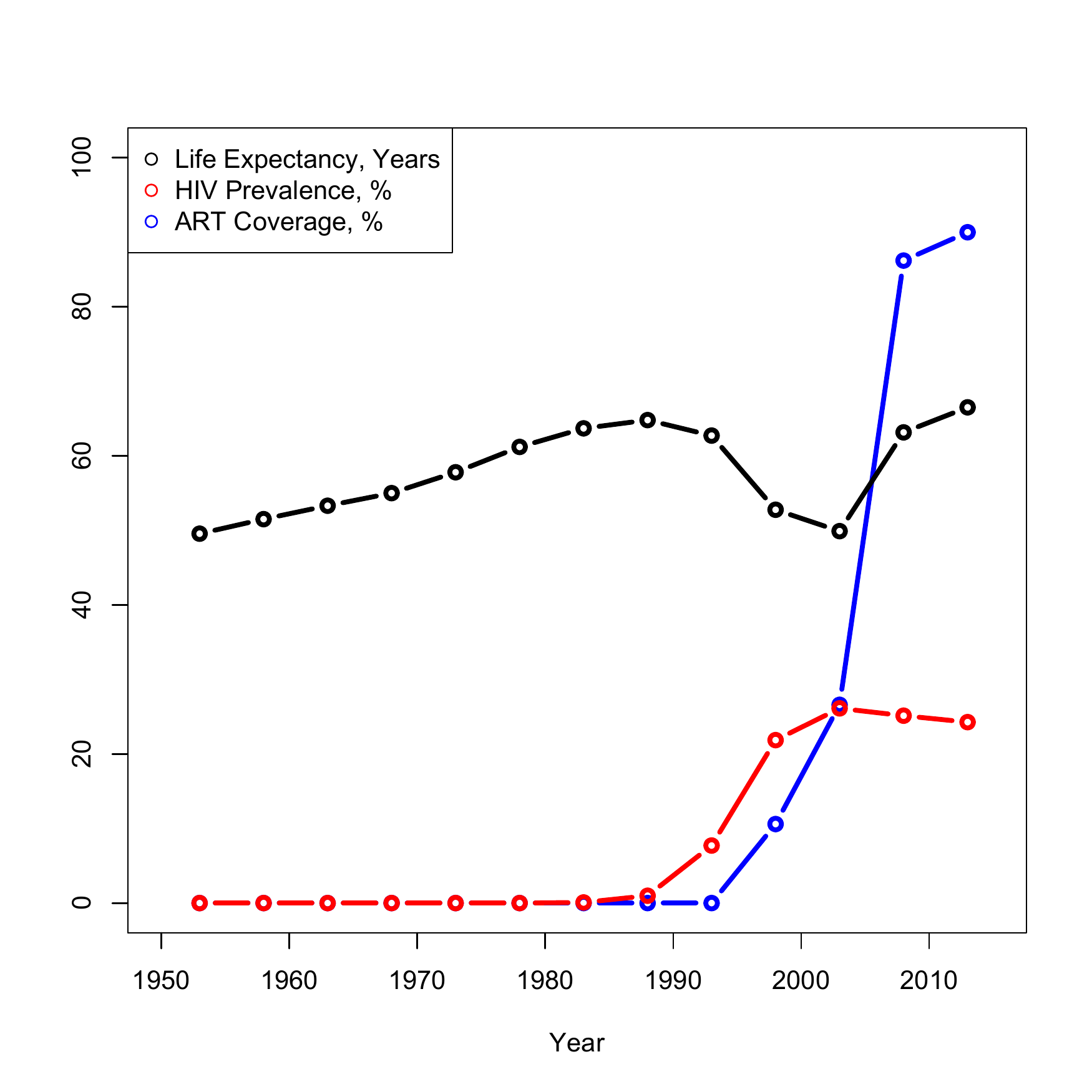}
\caption{Life expectancy at birth (black), HIV prevalence (red) and ART coverage (blue) for Botswana from 1950-1955 to 2005-2015.}
\label{fig:Botse0HIV}
\end{figure}

Figure \ref{fig:Botse0HIV} shows the rise of the HIV epidemic and the corresponding evolution of life expectancy at birth in Botswana. There was a sharp lowering of life expectancy with the rise of the epidemic, and then a rapid recovery to pre-epidemic levels following the widespread introduction of ART.

To incorporate covariate information into a probabilistic projection model, we must also have a method for projecting the covariate of interest into the future. UNAIDS developed the Spectrum/EPP methodology for projecting HIV prevalence and demographic measures, including life expectancy at birth, while accounting for HIV prevalence and ART coverage among other things 
\citep{stover:2012:spectrumEPP, stanecki:2012:HIVproj, spectrum:2014:software}.
The method is quite complicated and requires fine-grained data on a number of demographic and health measures for each country. It is recommended for reconstructing the HIV epidemic, including the time of onset, in a particular country and for projecting the epidemic up to five years into the future. It is not designed for the longer term projections that are important for long-term population projections \citep[p. 9]{spectrum:2014:guide}. 

We use a version of the EPP package for \texttt{R} for projections of HIV prevalence to 2100 \citep{brown:2010:EPP}. We develop a simpler model for projecting life expectancy at birth while accounting for HIV prevalence and ART coverage that is more practical for long term projections. 

\section{Methodology}
\label{sec:Methods}

\subsection{Data}
\label{subsec:data}

We use estimates of female life expectancy at birth from the United Nations \textit{World Population Prospects} (WPP) 2015 Revision \citep{un:WPP2015} for 201 countries. The UN produces estimates of period life expectancy at birth and age-specific mortality rates by five-year periods and five-year age groups; these are updated every two years. There are estimates for each country of the world for each five year period from 1950 to 2015. We do not use life expectancy inputs for Cambodia and Rwanda from the time periods of the genocides in these countries.  To fit the model, we use UNAIDS estimates of past HIV prevalence and ART coverage for 40 countries with generalized epidemics. We use 1000 trajectories of HIV prevalence, using the same assumptions as UNAIDS does in their projections. Additionally, we use a single deterministic trajectory of ART coverage from UNAIDS in the projection stage. We code HIV prevalence as zero for all countries not experiencing a generalized HIV epidemic.

\subsection{Review of joint probabilistic projections of male and female life expectancy}
\label{subsec:jointe0}

Our methodology builds on the Bayesian hierarchical model for 
probabilistic projection of female and male life expectancy used by the UN
\citep{raftery:2013:e0,raftery:2014:jointe0}, which proceeds as follows.
First, the Bayesian hierarchical model for female life expectancy is estimated using Markov chain Monte Carlo (MCMC), then probabilistic projections of female life expectancy are made from the present day to 2100. Projections of male expectancy are then made based on the projected values of female life expectancy \citep{raftery:2014:jointe0}.
The model provides a way for estimation and projection for one country to be improved using information from other countries. 

At the lowest (observation) level, 
the  Bayesian hierarchical model for female life expectancy at birth is 
\begin{eqnarray}
\Delta \ell_{c,t} \equiv  \ell_{c,t+1} - \ell_{c,t} & = &  
g(\ell_{c,t} \vert \theta^{(c)}) + \varepsilon_{c,t+1},
\label{eq:e0noHIV} \\
\varepsilon_{c,t} & \sim&  N(0, (\omega f(\ell_{c,t}))^2),
\label{eq:epsct}
\end{eqnarray}
where $\ell_{c,t}$ is the female life expectancy at birth for country $c$ in time period $t$, $g( \cdot \vert \theta^{(c)})$ is the expected five-year gain in life expectancy, modeled as a double logistic function of current life expectancy and governed by country-specific parameters, $\theta^{(c)}$, $\varepsilon_{c,t+1}$ is a random perturbation around the expected gain, and $f(\ell_{c,t})$ is a smooth function of life expectancy. 
The double logistic function for country $c$ is
\begin{equation} \label{eq:doublelogistic}
g(\ell_{c,t} \vert \theta^{(c)}) = \dfrac{k^c}{1+ \exp \left(-\frac{A_1}{\Delta_2^c}(\ell_{c,t} - \Delta_1^c - A_2\Delta_2^c)\right)}
+ \dfrac{z^c - k^c}{1 + \exp \left(-\frac{A_1}{\Delta_4^c}(\ell_{c,t} - \sum_{i=1}^3\Delta_i^c - A_2 \Delta_4^c)\right)},
\end{equation}
where $\theta^{(c)} = (\Delta_1^c, \Delta_1^c, \Delta_1^c, \Delta_1^c, k^c, z^c)$ and $A_1$ and $A_2$ are constants. The parameter $z^c$ is the expected country-specific asymptotic five-year gain in life expectancy. The other parameters govern the maximum value and the pace of rise and fall of expected five-year gains in life expectancy.  

At the second level of the model, the country-specific parameters 
$\theta^{(c)}$ are assumed to be drawn from the following world distribution:
\begin{center}
$\begin{array}{lr}
\Delta_i^c \vert \sigma_{\Delta_i} \stackrel{\text{iid}}{\sim} \text{Normal}_{[0,100]}( \Delta_i, \sigma_{\Delta_i}^2), & i = 1,\dots,4,\\
k^c \vert \sigma_k \stackrel{\text{iid}}{\sim} \text{Normal}_{[0,10]}(k, \sigma_k^2) &  \\
z^c \vert \sigma_z \stackrel{\text{iid}}{\sim} \text{Normal}_{[0,0.653]}(z,\sigma^2_z). & \\
\end{array}$
\end{center}
At the third, top level of the model, prior distributions are specified for
the world parameters  $\theta = (\Delta_1, \Delta_2, \Delta_3, \Delta_4, k, z, 
\omega)$.
\vspace{0.1in}

The Bayesian hierarchical model is estimated using MCMC via Metropolis-Hastings, Gibbs sampling and slice sampling steps, yielding a joint posterior distribution of all model parameters \citep{raftery:2013:e0}. The smooth function $f(\ell_{c,t})$ specifying the variance of the perturbations is estimated separately and is treated as known in the MCMC algorithm.

Once the model has been estimated, projections of life expectancy are made based on each posterior sample of $\theta^{(c)}$ and a random perturbation, $\varepsilon_{c,t+1}$, drawn from a $N(0, (\omega f(\ell_{c,t}))^2)$ distribution, where $\omega$ is drawn from the posterior distribution. 
After female projections of life expectancy are made, projections of male life expectancy, $\ell_{c,t}^m$, are made by modeling the gap between the two \citep{raftery:2014:jointe0}.

\subsection{Probabilistic Projections of Life Expectancy Accounting for HIV Prevalence}
\label{subsec:e0HIV}

 We expand the BHM to account for generalized HIV/AIDS epidemics by adding
a covariate to the observation level of the model. The covariate is based on
$HIVnonART_{c,t} = HnA_{c,t}$, defined as follows.
Let $HIV_{c,t}$ and $ART_{c,t}$ be the HIV prevalence and ART coverage in percent of country $c$ at time period $t$, respectively. Then $HnA_{c,t} = HIV_{c,t} \times (100 - ART_{c,t})$; it can be viewed as approximating the percentage of the population who are infected but do not receive ART therapy. 
The covariate we found to best predict change in life expectancy 
was the change in this quantity, namely 
$\Delta HnA_{c,t-1} = HnA_{c,t} - HnA_{c,t-1}$.
Our expanded observation equation is then
\begin{equation} \label{eq:e0HnA}
\Delta \ell_{c,t} = g(\ell_{c,t} \vert \theta^{(c)}) + \beta_{HnA} \Delta HnA_{c,t-1} + \varepsilon_{c,t+1}.
\end{equation}

The parameter $\beta_{HnA}$ is constant across countries and is estimated by MCMC along with the other parameters of the Bayesian hierarchical model. It has a diffuse prior distribution, chosen to be spread out enough that it has little impact on the final inference. Specifically, the prior distribution of $\beta_{HnA}$  is $N\left( 0, 0.25 \times \dfrac{Var(\Delta \ell_{c,t})}{Var(\Delta HnA_{c,t})}\right)$, where the prior variance is determined by the sample variances of observed changes in life expectancy and observed changes in $HnA$. The posterior distribution of $\beta_{HnA}$ is estimated with the other parameters in the MCMC via Gibbs sampling updates. 

After estimation, we project female life expectancy in the same manner as outlined in Section~\ref{subsec:jointe0}. However, we make a projection based on each posterior sample of $(\theta^{(c)}, \beta_{HnA})$ and a random perturbation. We account for uncertainty in the HIV trajectories by using 1000 yearly trajectories of HIV projections from EPP \citep{brown:2010:EPP}. For each country $c$ and year $t$, we find the median, $z_{t,c}$, of projected adult HIV prevalence output from EPP. We use a single UNAIDS deterministic projection to 2100 as a baseline reference, and we construct 1000 trajectories from the single UNAIDS trajectory by using 1000 multipliers of the form $\dfrac{z^k_{t,c}}{z_{t,c}}$, at each time point $t$ for $k = 1, \dots, 1000$, where $z^k_{t,c}$ is prevalence at time $t$ in country $c$ in the $k$ simulated trajectory. 
Thus the UNAIDS deterministic trajectory serves as the median trajectory of HIV prevalence to 2100, and the EPP trajectories determine the uncertainty. 
We construct 5-year averages from the yearly trajectories to be used in the projection stage. From these, we use a single deterministic trajectory of ART coverage to compute 1000 trajectories of $\Delta HnA_{c,t}$ for all countries. We sample one out of 1000 trajectories of $\Delta HnA_{c,t}$ with equal probability to be used in the projection stage. \par

For the country of Liberia, the prevalence is projected to be so low in the future (nearly 0) that the multipliers are unrealistically large. We therefore treat it slightly differently. For this country we calculate $z^k_{t,c}-z_{t,c}$ for each time point $t$. We then add this distance to the UNAIDS trajectory to yield 1,000 trajectories with the UNAIDS trajectory as the median and borrow the uncertainty from the EPP trajectories. \par

The methods of \citet{raftery:2013:e0} and \citet{raftery:2014:jointe0} did not use the generalized HIV epidemic countries in model estimation. 
By contrast, our estimation of the BHM does include these countries. 
The covariate values are set to zero for non-epidemic countries. 
Thus, the estimation of country-specific parameters and the projection of life expectancy for non-epidemic countries changes negligibly; we are fitting the model in \eqref{eq:e0noHIV} for these countries. For epidemic countries, the model in \eqref{eq:e0HnA} allows us to adjust for the effects of HIV on life expectancy in the linear term and to interpret $g(\cdot \vert \theta^{(c)})$ as the expected five-year gain in life expectancy in the absence of the epidemic. 
Though high AIDS prevalence takes a big toll on a country's life expectancy at birth in the absence of ART, ART extends an infected person's life substantially. Several epidemiological case studies show that patients have nearly normal life expectancy when treated with ART \citep{mills:2011:ARTe0, johnson:2013:ARTe0}. In a country where ART coverage is high, a generalized HIV epidemic affects life expectancy like a chronic disease \citep{deeks:2013:HIVchronic}. \par

In a manner similar to \citet{raftery:2013:e0}, the distribution of random perturbations in the projection stage is $\varepsilon_{c,t+1} \sim N(0, (\omega \times f(\ell_{c,t,i})^2)$, where $\omega$ is a model parameter, $f(\ell_{c,t,i})$ is a smooth function and $i$ is an indicator for generalized HIV epidemic. To estimate $f(\ell_{c,t,i})$, we fit the model in \eqref{eq:e0HnA} using the same function $f(\ell_{c,t})$ as used by \citet{raftery:2013:e0}. Then, using mean posterior estimates of $g(\ell_{c,t} \vert \theta^{(c)})$, we projected life expectancy forward from 1950-1955 to the 2010-2015 period using only the mean model in \eqref{eq:e0HnA} with no random perturbations. We then calculated absolute residuals for these projections.\par
\begin{figure}[h]
\centering
\includegraphics[scale=0.55]{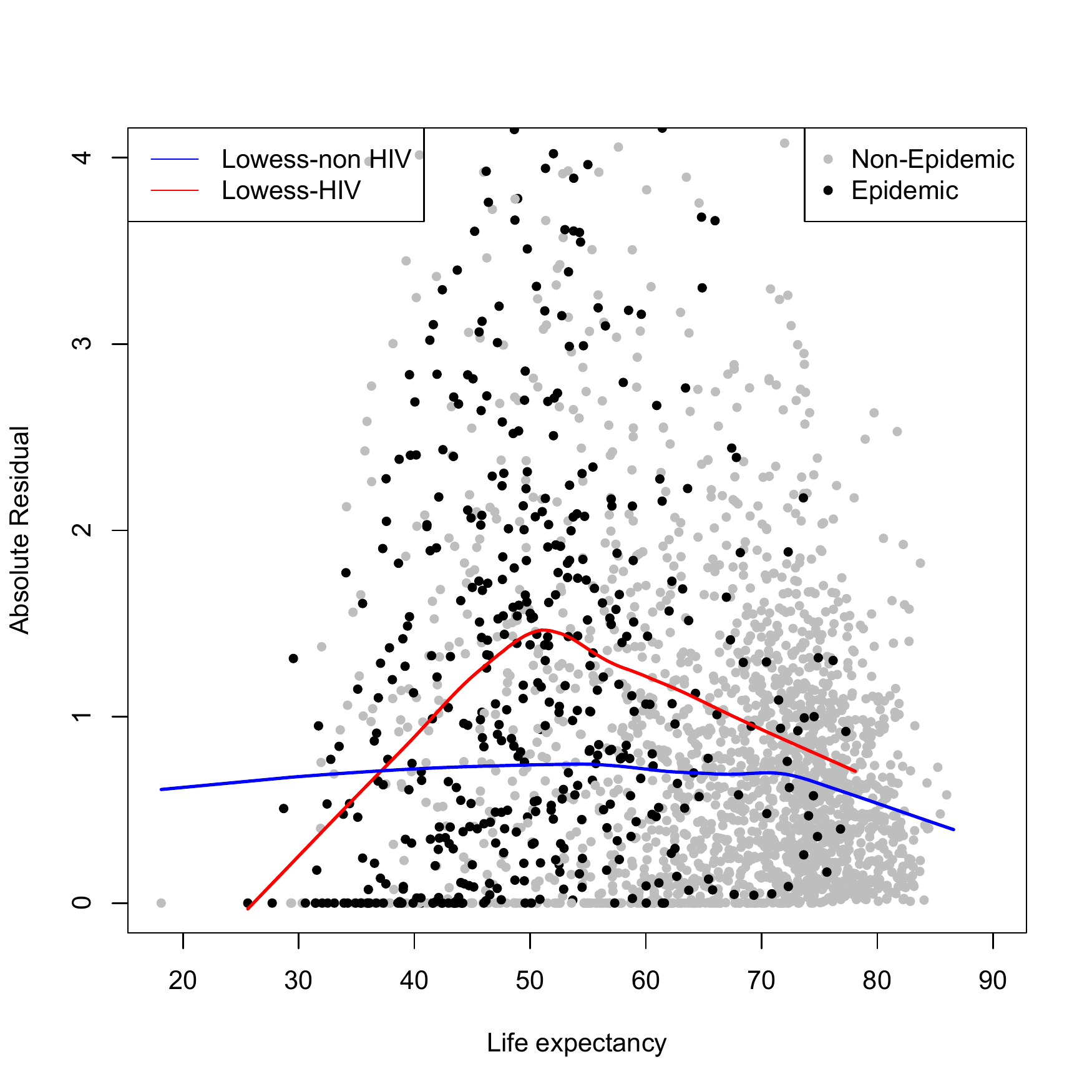}
\caption{The black dots represent absolute residuals for epidemic countries; the grey dots represent absolute residuals for non-epidemic countries. The blue line is the loess fit to the non-epidemic residuals. The red line is the loess fit to the epidemic residuals.}
\label{fig:loess}
\end{figure}

We fit loess curves to the absolute residuals for non-epidemic countries and for epidemic countries separately. These curves can be seen in Figure~\ref{fig:loess}. The black dots represent the absolute residuals from HIV countries, and the red loess curve is fit to these points. The grey dots represent the absolute residuals from non-HIV countries, and the blue curve is fit to these points. Here one can see that the HIV countries have more variability than the non-HIV countries. This variability is disseminated into the future for countries currently experiencing a generalized epidemic. For non-HIV countries, $f(\ell_{c,t,i=nonHIV})$ is the blue curve in Figure~\ref{fig:loess}. For HIV countries, $f(\ell_{c,t,i=HIV})$ is the maximum of the blue and red curves up to the highest observed life expectancy for an HIV country to date, namely 78.1. For projected female life expectancies above 78.1 years, we use the blue curve plus a constant that is the vertical difference between the red and blue curves at 78.1 years.

\par

\subsection{Model Validation}
\label{subsec:Validation}

\begin{table}[ht]
\centering
\caption{Predictive Validation Results for Female Life Expectancy. 
The first column represents the set of countries used in the subsequent calibration calculations. The second column represents the time period of data used to fit the model, and the third column represents the time periods used in validation. The fourth column represents the number of countries used in validation. In the fifth column, ``No Covariates" represents the model in \eqref{eq:e0noHIV} and ``$\Delta HnA$" represents the model in \eqref{eq:e0HnA}. The sixth column contains the MAE as defined in Section~\ref{subsec:Validation}. The seventh and eighth columns contain coverage metrics for the 80\% and 95\% predictive intervals respectively.}
\vspace{0.2in}
\begin{tabular}{|c|c|c|c|c|c|cc|}
\hline
\multicolumn{1}{|c|}{Countries} & \multicolumn{1}{c|}{Training} &\multicolumn{1}{c|}{Test} &\multicolumn{1}{c|}{$n$} &\multicolumn{1}{c|}{Model} & \multicolumn{1}{c|}{MAE} & \multicolumn{2}{c|}{Coverage}  \\ \cline{7-8}
 & Period & Period & & & & 80\% & 95\%   \\
\hline
\multirow{8}{*}{HIV} & \multirow{2}{*}{1950-2005} & \multirow{2}{*}{2005-2015} & \multirow{2}{*}{69} & No Covariates & 3.47 & 0.49 & 0.58  \\
& & & & $\Delta HnA$ & 2.29 & 0.71 & 0.87 \\ \cline{2-8}
& \multirow{2}{*}{1950-2005} & \multirow{2}{*}{2005-2010} & \multirow{2}{*}{40} & No Covariates & 2.40  & 0.53 & 0.63 \\
& & & & $\Delta HnA$ & 2.22 & 0.68 & 0.83 \\ \cline{2-8}
& \multirow{2}{*}{1950-2005} & \multirow{2}{*}{2010-2015} & \multirow{2}{*}{29} & No Covariates & 4.95 & 0.45 & 0.52  \\
& & & & $\Delta HnA$  & 2.39 & 0.76 & 0.93\\ \cline{2-8}
& \multirow{2}{*}{1950-2010} & \multirow{2}{*}{2010-2015} & \multirow{2}{*}{29} & No Covariates & 2.24  & 0.59 & 0.62\\
& & &  & $\Delta HnA$  & 1.74 & 0.83 & 0.97 \\ \cline{2-8}
\hline
\multirow{8}{*}{All} & \multirow{2}{*}{1950-2005} & \multirow{2}{*}{2005-2015} & \multirow{2}{*}{390} & No Covariates & 1.15 & 0.83 & 0.91 \\
& & & & $\Delta HnA$ & 0.94 & 0.88 & 0.96 \\ \cline{2-8}
& \multirow{2}{*}{1950-2005} & \multirow{2}{*}{2005-2010} & \multirow{2}{*}{201} & No Covariates & 0.84 & 0.85 & 0.92 \\
& & & & $\Delta HnA$ & 0.81 & 0.90 & 0.96\\ \cline{2-8}
& \multirow{2}{*}{1950-2005} & \multirow{2}{*}{2010-2015} & \multirow{2}{*}{189} & No Covariates & 1.49 & 0.80 & 0.91 \\
& & & & $\Delta HnA$ & 1.09 & 0.86 & 0.97\\ \cline{2-8}
& \multirow{2}{*}{1950-2010} & \multirow{2}{*}{2010-2015} & \multirow{2}{*}{189} & No Covariates & 0.74 & 0.84 & 0.91  \\
& & & & $\Delta HnA$ & 0.66 & 0.89 & 0.97  \\ \cline{2-8}
\hline
\end{tabular}
\label{tab:modelvalid}
\end{table}

We performed predictive out of sample validation to assess our model. First, we fit the model in \eqref{eq:e0HnA} using data from 1950-1955 up to 2000-2005 and projected female life expectancy for the time periods 2005-2010 and 2010-2015. 
We also fit the model in \eqref{eq:e0HnA} using data from 1950-1955 up to 2005-2010, and projected female life expectancy for the time period 2010-2015. 

Table~\ref{tab:modelvalid} presents our results. The first column designates the set of countries for which the metrics have been calculated. Note that though the calibration of predictive intervals is designed to be nominal for ``All" countries, we include the subset of HIV countries for more detail.  The second and third columns reflect the period of data used to train the model and validate the model, respectively. The fourth column contains the number of countries for which the subsequent calibration metrics are calculated. 

There were 12 countries for which no new data became available between the
publication of the 2012 UN estimates in WPP 2012 \citep{un:WPP2012} and the 2015
UN estimates in WPP 2015 \citep{un:WPP2015}.
Hence, the WPP 2015 life expectancy estimate for the time period 2010-2015 is actually the projection of life expectancy for this period from the WPP 2012. As such, there is no ``observed" life expectancy for these countries for the time period 2010-2015, and so we excluded these countries in this time period from the validation exercise. In the fifth column, ``No Covariates" refers to the model in \eqref{eq:e0noHIV} and ``$\Delta HnA$" refers to the model in \eqref{eq:e0HnA}. 

The last three columns contain our metrics.
The mean absolute error (MAE) is calculated as 
\begin{equation}
\frac{1}{n} \sum_{c \in {\cal C}} \sum_{t \in {\cal T}}
\vert \hat{\ell}_{c,t} - \ell_{c,t} \vert,
\label{eq-MAE}
\end{equation}
where $\hat{\ell}_{c,t}$ is the median projection of female life expectancy for country $c$ in time period $t$.
In (\ref{eq-MAE}), ${\cal C}$ is the set of countries involved in calculating
the MAE (either the HIV countries or all countries), 
${\cal T}$ is the set of five-year time periods involved as shown in the third
column, and $n$ is the number of country-time period combinations as
shown in the fourth column.
The last two columns show the proportion of countries whose 80\% and 95\% posterior predictive intervals contain the observed life expectancy in the validation period of interest.\par

In all the out of sample scenarios, we saw substantial improvements in coverage for HIV countries after accounting for HIV prevalence and ART coverage. We broke down the two-period out of sample exercise into the two projection periods to get more detailed information about the HIV countries.  In 2005-2010, the model with no covariates missed 15 out of 40 HIV countries at the 95\% level. When we accounted for HIV prevalence and ART coverage, the number of HIV countries missed went down by over half to only 7 at the 95\% level. 
For 2010-2015, the model with no covariates missed 14 out of 29 HIV countries at the 95\% level, and accounting for HIV prevalence and ART coverage
reduced this to only 2 HIV countries in the time period 2010-2015. 
In the leave-two-time-periods-out validation exercise, we saw a decrease in MAE for HIV countries in every case, while the MAE remained unchanged for non-HIV countries. For the non-HIV countries, the addition of the covariate in the model in \eqref{eq:e0HnA} changed coverage negligibly.

When fitting the model using data from 1950-1955 up to 2005-2010 and projecting female life expectancy for 2010-2015, we also saw improvements. Our coverage was closer to nominal for HIV countries after accounting for HIV prevalence and ART coverage. We missed 11 HIV countries out of 29 in the model with no covariates, but only one HIV country after accounting for the HIV epidemic. The MAE also decreased when accounting for HIV prevalence and ART coverage.\par

Predictive validation results for male life expectancy are shown in
Table \ref{tab:modelvalidM} in Appendix B, and the conclusions are broadly
similar.

The current method used by the UN to project life expectancy in the presence of the HIV epidemic is the Spectrum/EPP package \citep{spectrum:2014:software, stanecki:2012:HIVproj, stover:2012:spectrumEPP}. 
However, Spectrum is a complicated model with heavy data demands and is intended only for short-term projections up to five years into the future. 
An important question is therefore whether our simpler method can produce short-term projections similar to those of the more complex Spectrum method.

\begin{figure}
\centering
\includegraphics[scale=0.45]{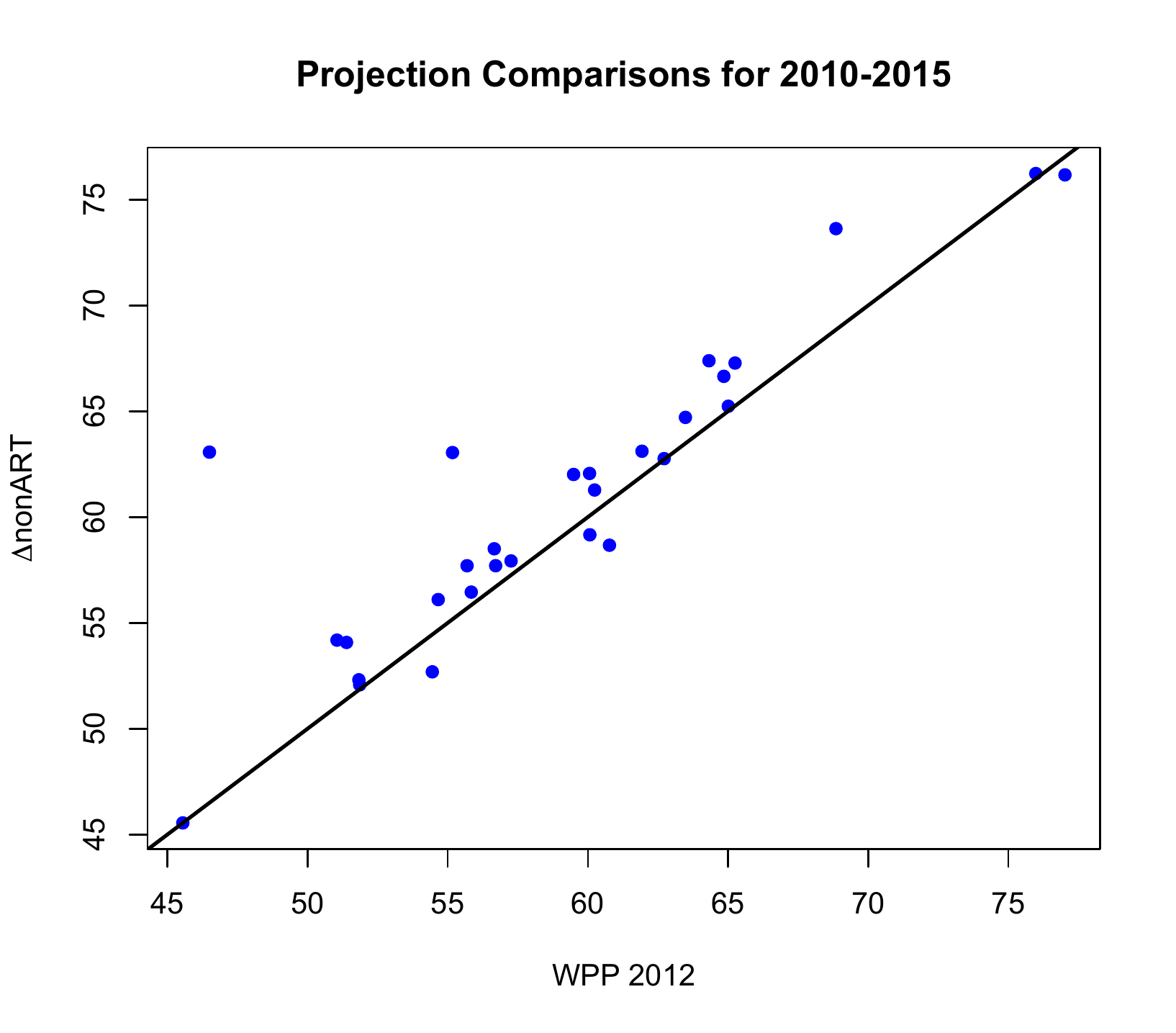}
\includegraphics[scale=0.45]{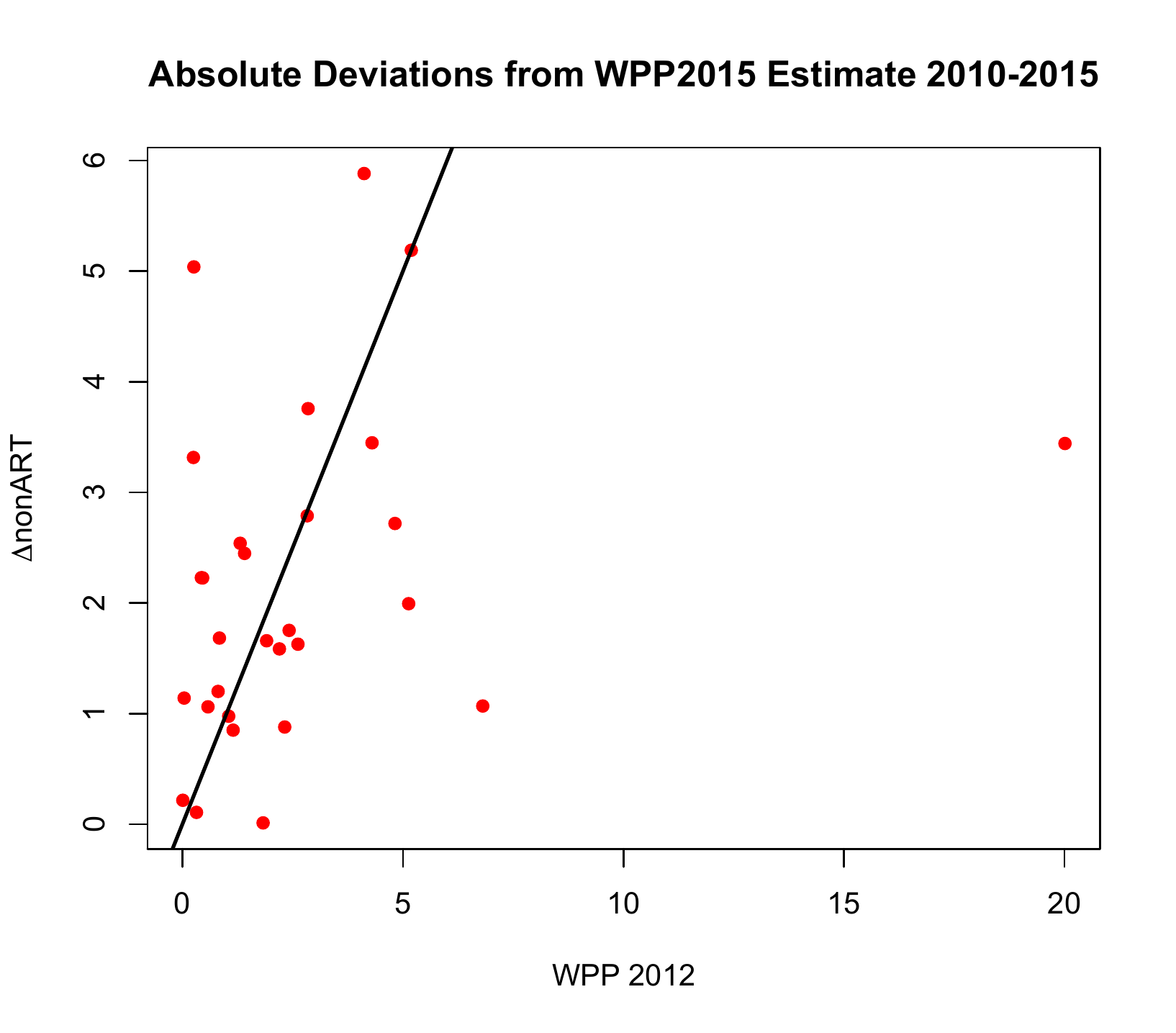}
\caption{Comparison between short-term projections in WPP 2012 using Spectrum, and our simpler method.
On the $x$-axis are projections of female life expectancy in 2010-2015 produced using Spectrum and published in WPP 2012. On the $y$-axis are projections of female life expectancy in 2010-2015 produced by fitting the model in \eqref{eq:e0HnA} with WPP2012 data up to 2005-2010. The projections are mostly similar and remain close to the $y=x$ line.}
\label{fig:spectrumcompare}
\end{figure}

To answer this, we fit our model in \eqref{eq:e0HnA} with WPP 2012 estimates of female life expectancy from 1950-1955 up to 2005-2010 \citep{un:WPP2012}. Then we projected female life expectancy to 2010-2015. We compared the projections from our simpler model designed to make long-term projections to the WPP 2012 projection for 2010-2015 made using Spectrum. In the left panel of Figure~\ref{fig:spectrumcompare}, we see that the five-year projections from our simpler method are similar to the projections from the more complicated Spectrum. In fact, the correlation between the projections from our proposed model and those published in WPP 2012 is 0.89.

 The right panel of Figure~\ref{fig:spectrumcompare} shows the absolute deviation from the WPP 2015 estimate of female life expectancy in 2010-2015 for our projections and the projections published in WPP 2012. The WPP 2012 projections have a mean absolute error of 2.70 years using the WPP 2015 estimate for comparison. The projections produced with the model in \eqref{eq:e0HnA} have a mean absolute error of 2.17 years. 

The large outlier in the right panel of Figure~\ref{fig:spectrumcompare} is the country of Botswana, and corresponds to the largest deviation seen in the left panel of Figure~\ref{fig:spectrumcompare}. Botswana has the highest HIV prevalence in the world, at 24.3\% in 2010-2015,  and has had a recent scale up of ART coverage. The boost in ART coverage yielded a rapid recovery in life expectancy of nearly 20 years from the WPP 2012 estimate for the 2005-2010 time period. Our model captures this large jump in life expectancy. In summary, our model produces similar projections to the current methodology designed for short-term projections, but using a much simpler model with smaller data requirements.

\section{Case Studies}
\label{sec:CaseStudies}
We now give specific results for five countries that illustrate specific
aspects of the method. Results for all countries we consider as having
generalized epidemics are given for female life expenctancy in Appendix A 
and for male life expectancy in Appendix B.

\subsection{Nigeria}

\begin{figure}
\centering
\includegraphics[height=7cm]{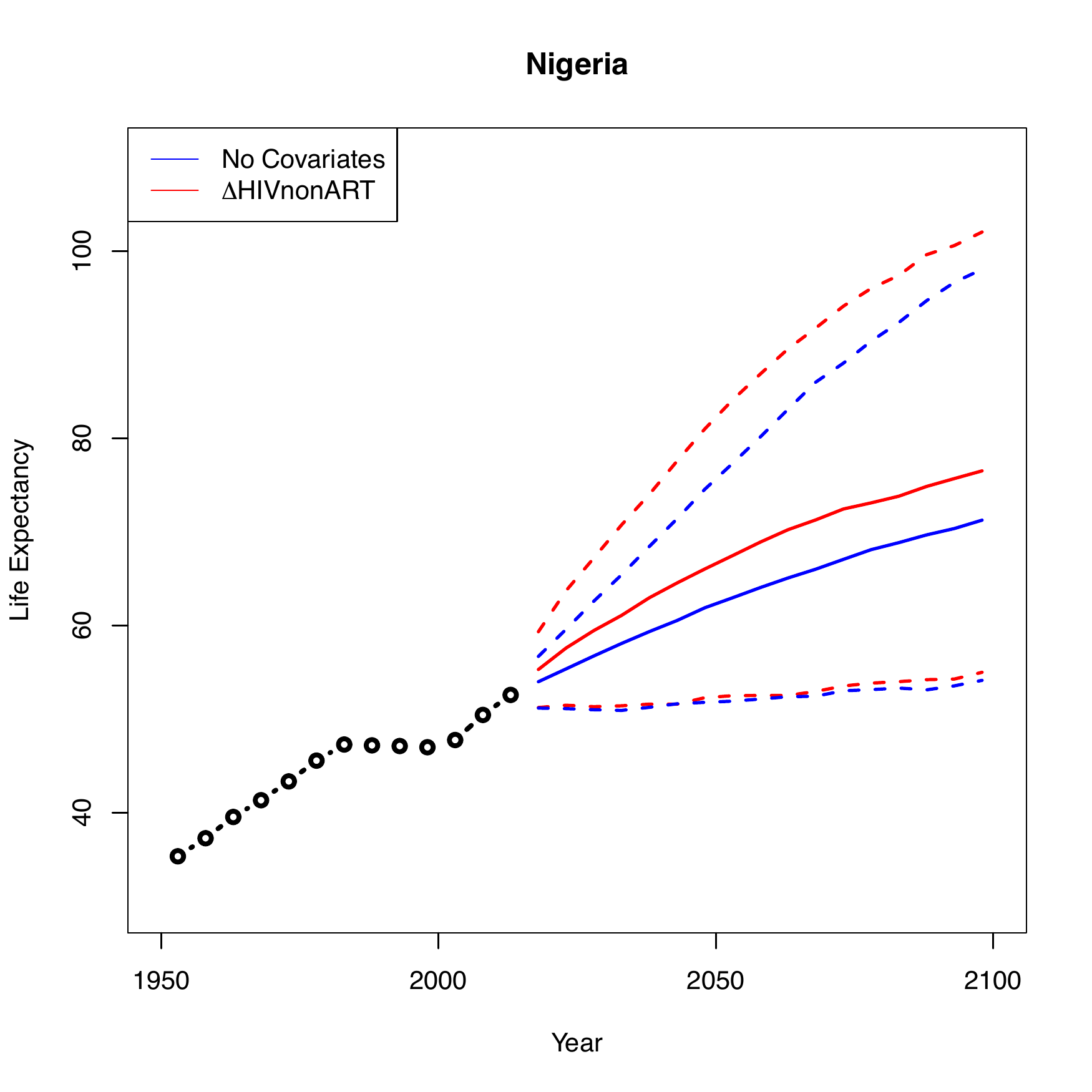}
\includegraphics[height=7cm]{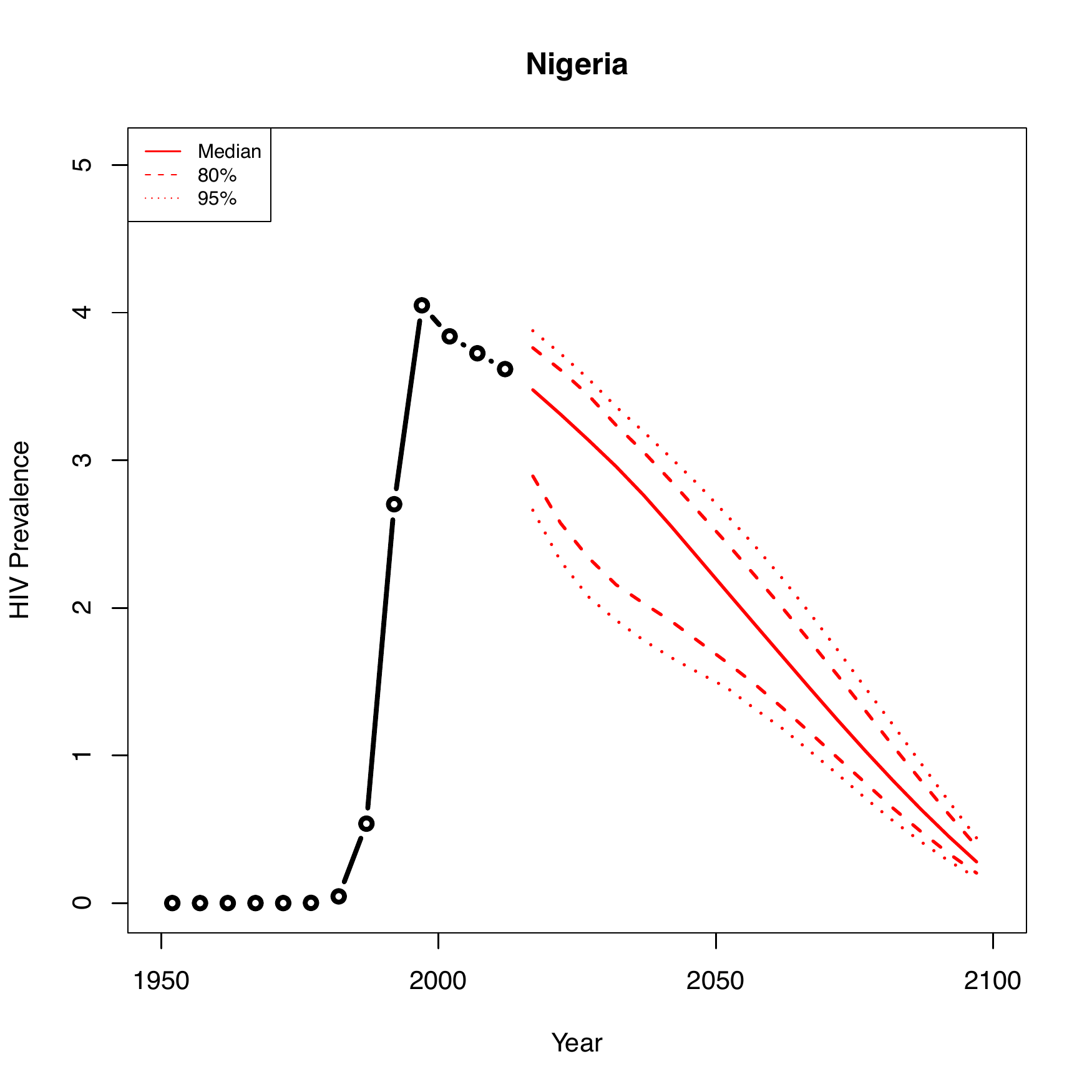}
\caption{The left panel shows projections of female life expectancy in Nigeria under the model in equation \eqref{eq:e0noHIV} in blue and equation \eqref{eq:e0HnA} in red. The solid lines represent medians, and the dashed lines are the 95\% intervals. After accounting for HIV prevalence and ART coverage, we see slightly more uncertainty about the future life expectancy in Nigeria and slightly higher median projections. The right panel shows the single trajectory of past estimates of HIV we use in model fitting in black. In red we have the median, 80\% interval and 90\% interval of probabilistic trajectories of HIV prevalence from EPP we use in our projections.}
\label{fig:Nigeria}
\end{figure}

Nigeria in West Africa is the most populous country in Africa. It has a relatively small epidemic, with prevalence of 3.6\% in 2010-2015. Figure~\ref{fig:Nigeria} shows a comparison between projections of life expectancy under the model \eqref{eq:e0noHIV} with no covariates in blue, and the model \eqref{eq:e0HnA} with the HIV covariate in red. 
The median projections of female life expectancy are higher than those 
projected when not accounting for the HIV/AIDS epidemic, and
accounting for the epidemic leads to more uncertainty about the future trajectory of female life expectancy in Nigeria.

\subsection{Kenya}
\begin{figure}
\centering
\includegraphics[height=7cm]{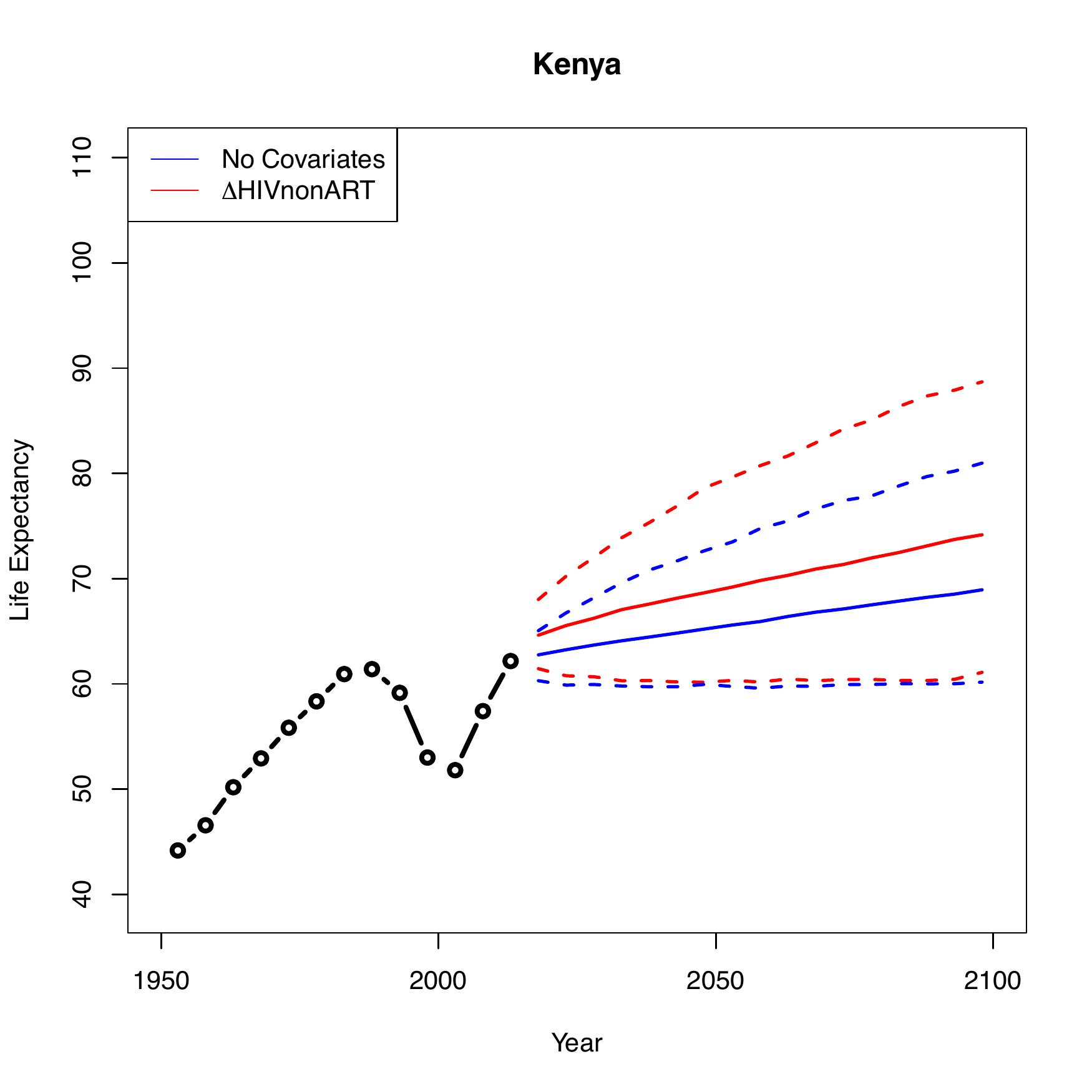}
\includegraphics[height=7cm]{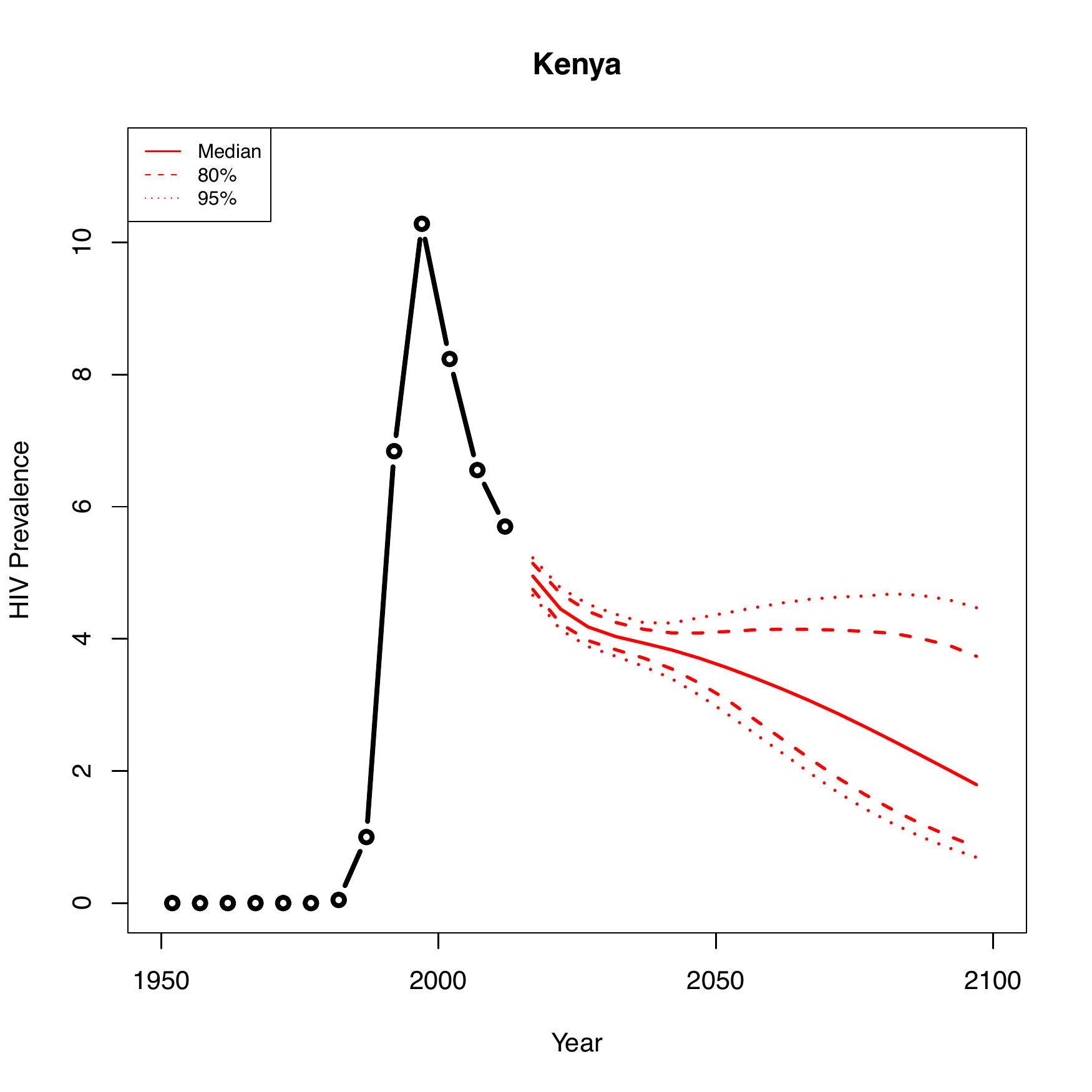}
\caption{The left panel shows projection of female life expectancy in Kenya under the model in equation \eqref{eq:e0noHIV} in blue and equation \eqref{eq:e0HnA} in red. The solid lines represent medians, and the dashed lines are the 95\% intervals. After accounting for HIV prevalence and ART coverage, we see a higher median projection of life expectancy with more uncertainty. The right panel shows the single trajectory of past estimates of HIV we use in model fitting in black. In red we have the median, 80\% interval and 90\% interval of probabilistic trajectories of HIV prevalence from EPP we use in our projections.}
\label{fig:Kenya}
\end{figure}

Kenya in East Africa has a medium-sized epidemic with HIV prevalence 5.7\% in 2010-2015.  Figure~\ref{fig:Kenya} shows that Kenya has already recovered to pre-epidemic life expectancy levels. After accounting for HIV prevalence and ART coverage, we project a slightly higher median female life expectancy to 2100 with more uncertainty at all time periods.

\subsection{South Africa}

\begin{figure}
\centering
\includegraphics[height=7cm]{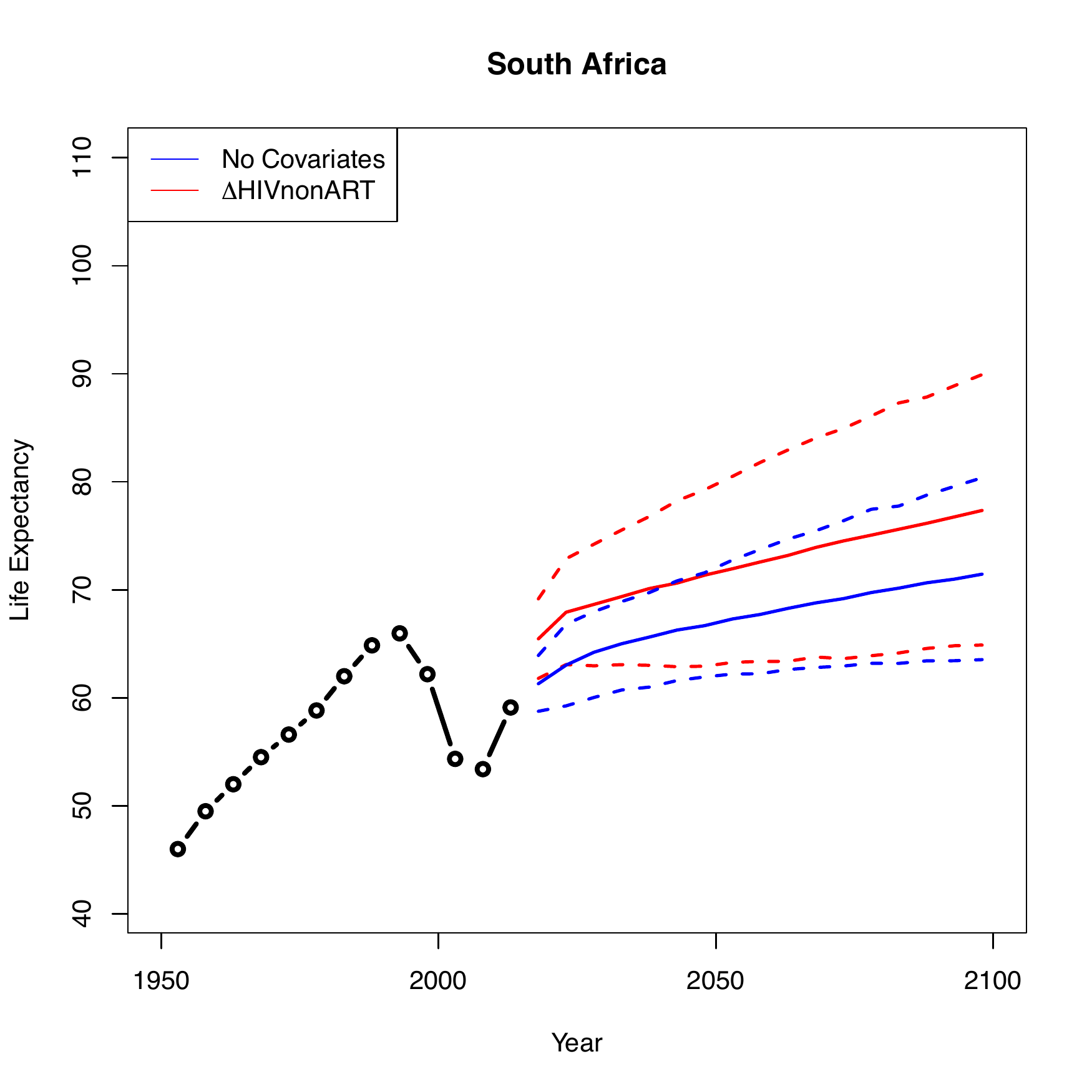}
\includegraphics[height=7cm]{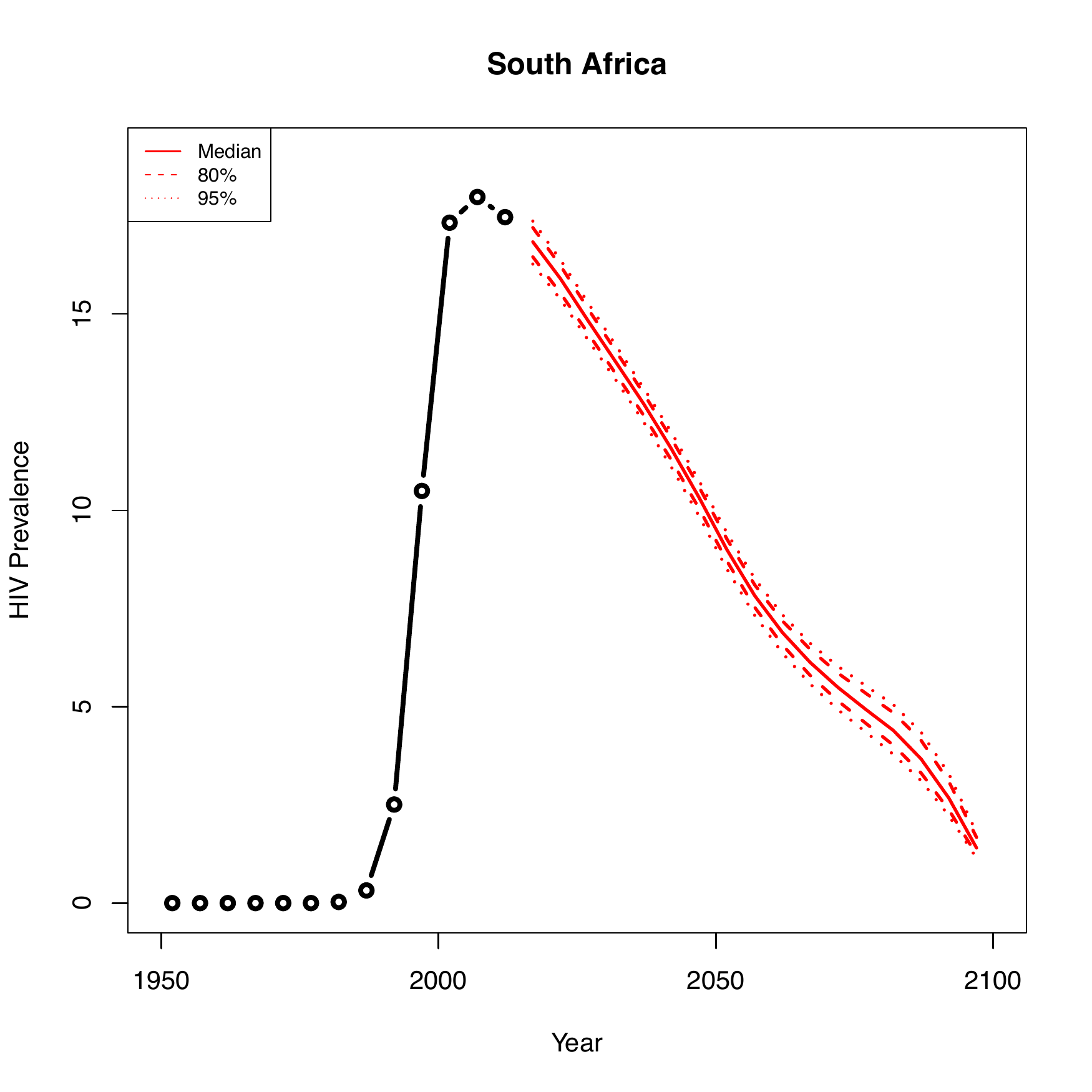}
\caption{The left panel shows projection of female life expectancy in South Africa under the model in equation \eqref{eq:e0noHIV} in blue and equation \eqref{eq:e0HnA} in red. The solid lines represent medians, and the dashed lines are the 95\% intervals. After accounting for HIV prevalence and ART coverage, we see an initial recovery of life expectancy to pre-epidemic levels followed by a steady rise through the end of the century. The right panel shows the single trajectory of past estimates of HIV we use in model fitting in black. In red we have the median, 80\% interval and 90\% interval of probabilistic trajectories of HIV prevalence from EPP we use in our projections.}
\label{fig:SouthAfrica}
\end{figure}

South Africa has the largest HIV/AIDS epidemic in the world in absolute numbers.
 The estimated prevalence in the 2010-2015 time period is 17.5\%. Figure~\ref{fig:SouthAfrica} shows a comparison between projections of life expectancy under the model in \eqref{eq:e0noHIV} in blue and the model in \eqref{eq:e0HnA} in red. Figure~\ref{fig:SouthAfrica} reflects the clear impact of ART coverage on recovery in life expectancy under a large epidemic. After accounting for HIV prevalence and ART coverage, we project an initial recovery to pre-epidemic life expectancy levels with a steady rise through the end of the century. When not accounting for the HIV epidemic and, particularly, ART coverage, the model projects median end of century life expectancy only slightly higher than South Africa's life expectancy before the HIV/AIDS epidemic. 
This is contrary to the epidemiological literature referenced in Section~\ref{subsec:e0HIV} that shows life expectancy recovers quickly after a scale-up of ART coverage.\par


As mentioned in Section~\ref{subsec:Validation}, there are a number of countries for which the UN did not have up-to-date life expectancy data at the time of publication of the WPP 2015, including South Africa. The UN estimates the female life expectancy for the period 2010-2015 as 59.1 years \citep{un:WPP2015}.
Statistics South Africa has published mid-year estimates of life expectancy 
for each calendar year (Statistics South Africa 2010, 2011, 2013, 2014, 2015),
\nocite{ssa:2010,ssa:2011,ssa:2013,ssa:2014,ssa:2015}
and averaging these gives an estimate for the five-year period 2010-2015
of 60.7 years.
When we fit our model with data up to 2005-2010 and project forward
five years to 2010-2015, our 95\% interval for 2010-2015 is (56.6, 65.1).
When we fit our model with data only up to 2000-2005 and project forward
ten years to 2010-2015, our 95\% interval for 2010-2015 is (61.3, 63.7).
In both cases, our interval captures the outcome, whether measured
by the UN or Statistics South Africa, and in particular the rapid increase
in life expectancy due to the widespread rollout of ART.

\subsection{Botswana}
Due to the early and rapid rise of HIV prevalence in Botswana, the ART scale-up was also quick. This led to the rapid recovery in life expectancy in the 2005-2010 time period. After accounting for the epidemic and antiretroviral therapy, we project a slightly slower rise in median life expectancy with more certainty than the model that does not account for the epidemic. This is in agreement with the epidemiological literature cited in Section~\ref{subsec:e0HIV} suggesting that HIV/AIDS will affect life expectancy as a chronic disease would after ART has become pervasive.

\begin{figure}
\centering
\includegraphics[height=7cm]{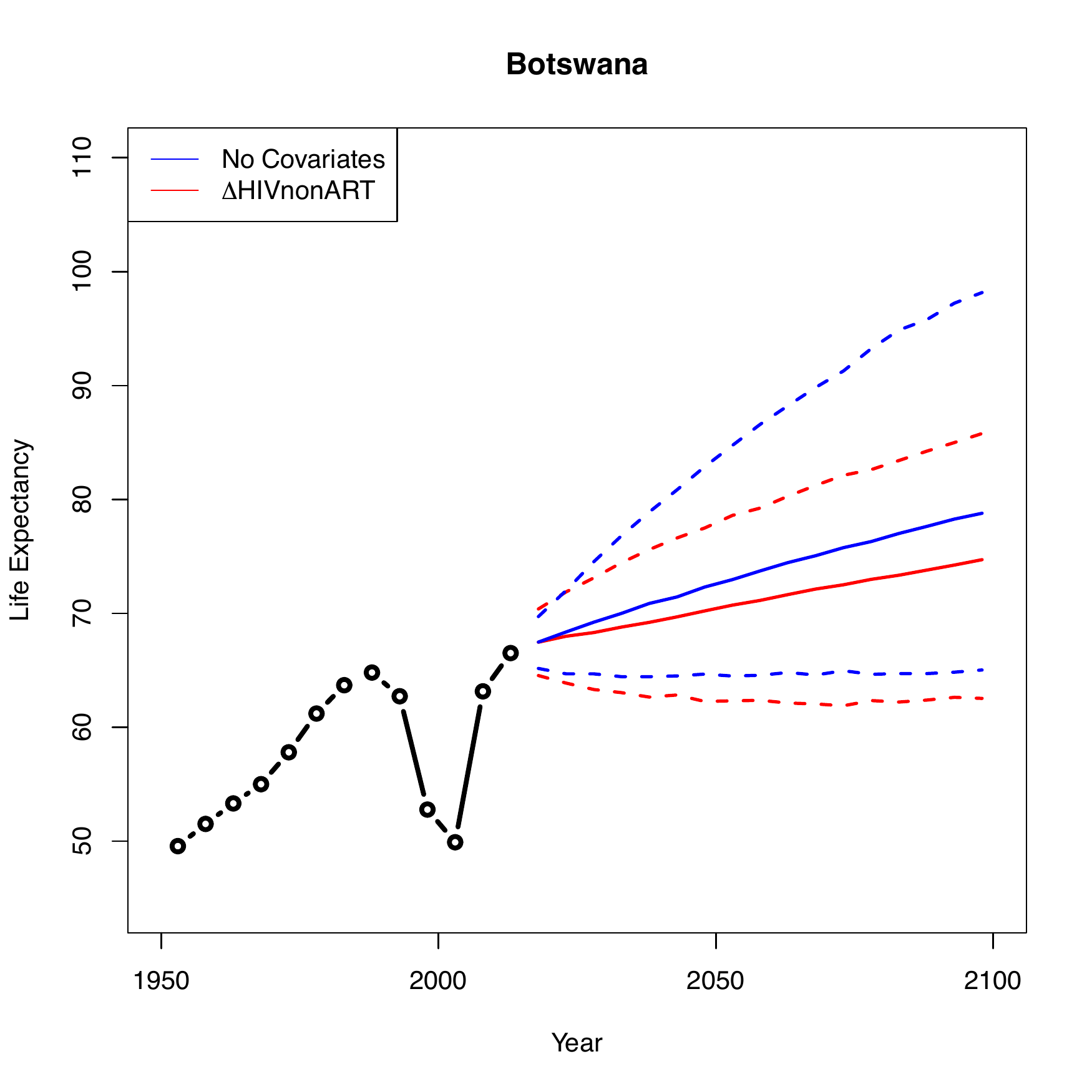}
\includegraphics[height=7cm]{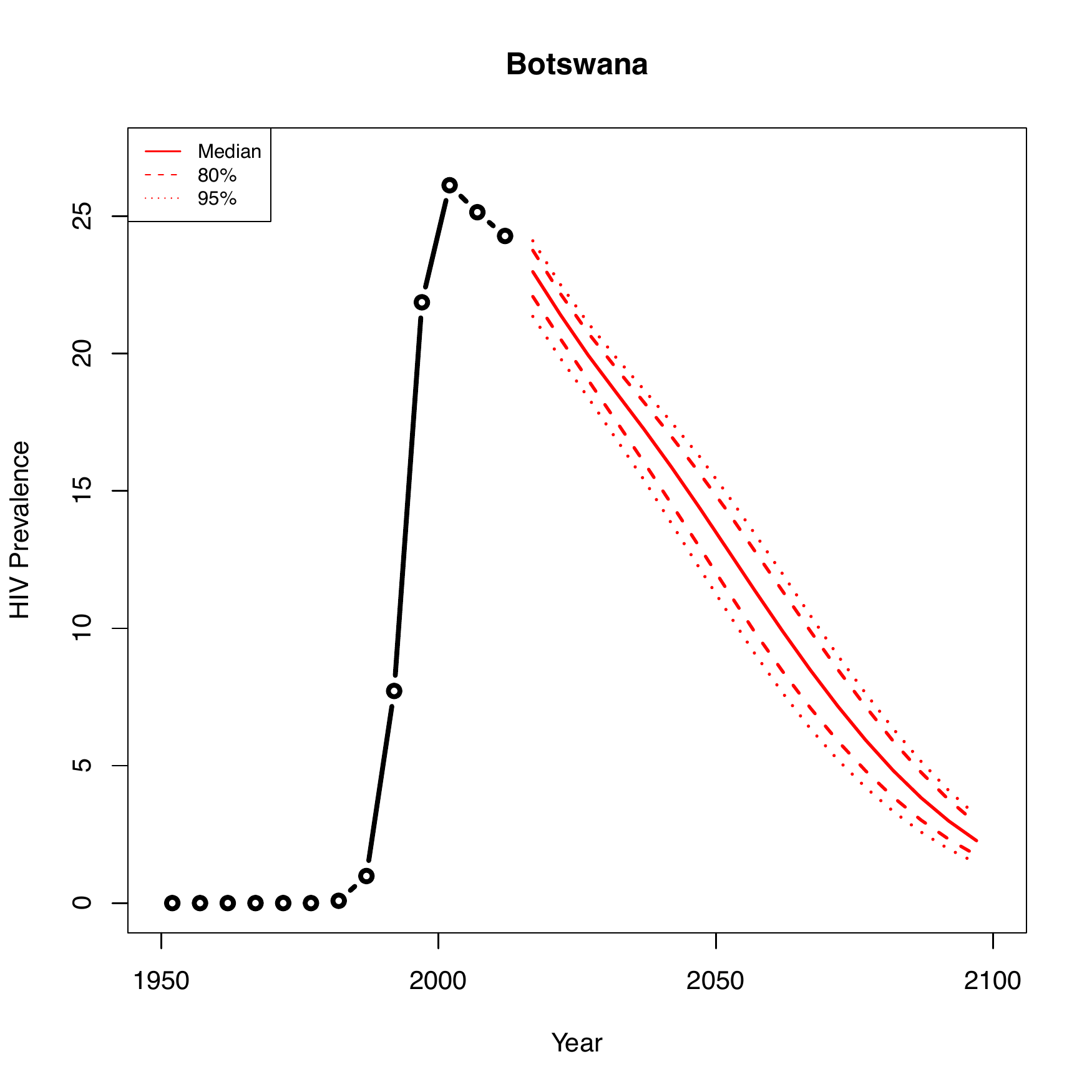}
\caption{The left panel shows projection of female life expectancy in Botswana under the model in equation \eqref{eq:e0noHIV} in blue and equation \eqref{eq:e0HnA} in red. The solid lines represent medians, and the dashed lines are the 95\% intervals. After accounting for HIV prevalence and ART coverage, we see a dampened, slow-rising median projection of life expectancy with less uncertainty.}
\end{figure}
\label{fig:Botswana}


\subsection{Germany}

\begin{figure}
\centering
\includegraphics[height=7cm]{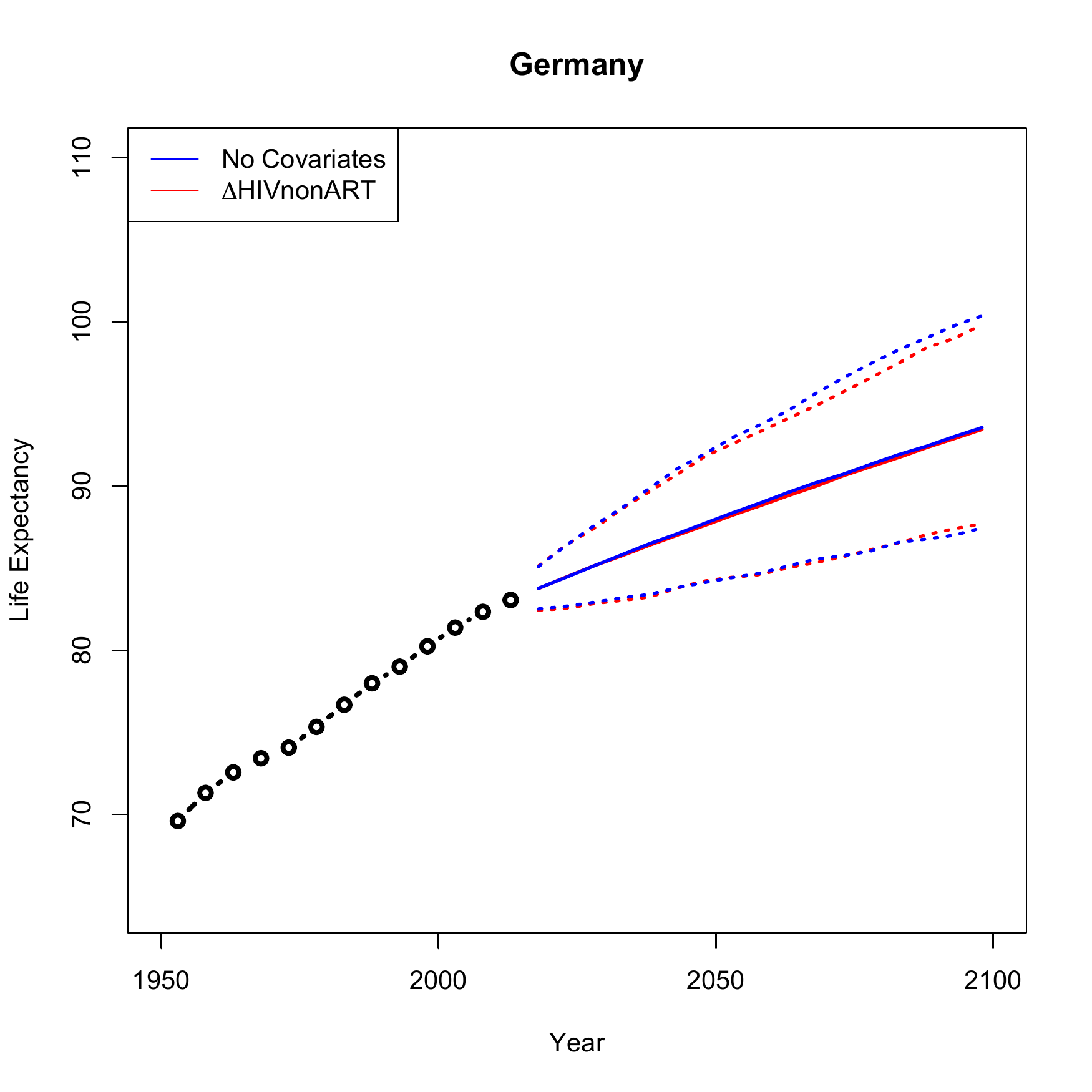}
\caption{This figure shows projection of female life expectancy in Germany under the model in equation \eqref{eq:e0noHIV} in black and equation \eqref{eq:e0HnA} in red. The medians are identical, and the predictive intervals differ negligibly.}
\label{fig:Germany} 
\end{figure}

Germany is an example of a country that does not have a generalized epidemic. As can be seen in Figure~\ref{fig:Germany}, projections of life expectancy under the model in \eqref{eq:e0noHIV} and the model in \eqref{eq:e0HnA} differ negligibly in both median and uncertainty.

\section{Discussion}
We have developed a probabilistic method for projecting life expectancy while accounting for generalized HIV/AIDS prevalence and antiretroviral therapy coverage. Our method has relatively modest data requirements. Through predictive validation we have shown that our method improves upon life expectancy projections for HIV/AIDS countries using the method in \citet{raftery:2013:e0}, while leaving projections for non-epidemic countries essentially unchanged. 
Our projections improve in terms of both the mean absolute error of point predictions and the calibration of predictive intervals. Our method produces similar short-term projections to the UNAIDS Spectrum/EPP package, with a simpler model that requires much less data. Moreover, the method can produce long-term projections out to 2100.\par
Our model reflects the literature consensus mentioned in Section~\ref{subsec:e0HIV} that HIV prevalence will have large impacts on life expectancy only in the absence of antiretroviral therapy. Once ART covers a large proportion of the infected population, there is a one-time gain in life expectancy to pre-epidemic levels and the effects will be modest afterwards.

One limitation of our method is the quality of the ART coverage data and projections. As ART coverage is relatively new and hard to measure, the data we have are noisy. Improvements in ART data quality would likely result in improvements in projections for the generalized HIV epidemic countries. Given good quality data, our method could also be extended to account for other covariates that explain changes in life expectancy. The data would need to be available for every country used in model fitting back to 1950. Methodology for projecting the covariates would also be required.\par

\section{Acknowledgements}
The project described was supported by grants R01 HD-054511 and R01 HD-070936 from the Eunice Kennedy Shriver National Institute of Child Health and Human
Development. The authors are grateful to Le Bao, Samuel Clark, Yanjun He and Hana \v{S}ev\v{c}\'{i}kov\'{a} for helpful discussions and sharing data and code.

\bibliography{JessRefs}

\begin{thebibliography}{}

\bibitem[\protect\citeauthoryear{Brown, Bao, Raftery, Salomon, Baggaley,
  Stover, and Gerland}{Brown et~al.}{2010}]{brown:2010:EPP}
Brown, T., L.~Bao, A.~E. Raftery, J.~A. Salomon, R.~F. Baggaley, J.~Stover, and
  P.~Gerland (2010).
\newblock Modeling {HIV} epidemics in the antiretroviral era: the {UNAIDS}
  {E}stimation and {P}rojection {P}ackage 2009.
\newblock {\em Sexually Transmitted Infections\/}~{\em 86}, 3--10.

\bibitem[\protect\citeauthoryear{Deeks, Lewin, and Havlir}{Deeks
  et~al.}{2013}]{deeks:2013:HIVchronic}
Deeks, S.~G., S.~R. Lewin, and D.~V. Havlir (2013).
\newblock The end of {AIDS}: {HIV} infection as a chronic disease.
\newblock {\em The Lancet\/}~{\em 382\/}(9903), 1525--1533.

\bibitem[\protect\citeauthoryear{{Futures Institute}}{{Futures
  Institute}}{2014}]{spectrum:2014:software}
{Futures Institute} (2014).
\newblock {Spectrum/EPP} 2014 version 5.06. (downloaded {June} 2014).
\newblock http://www.futuresinstitute.org/spectrum.aspx.

\bibitem[\protect\citeauthoryear{Girosi and King}{Girosi and
  King}{2008}]{girosi:2008:demforecasting}
Girosi, F. and G.~King (2008).
\newblock {\em Demographic Forecasting}.
\newblock Princeton University Press.

\bibitem[\protect\citeauthoryear{Johnson, Mossong, Dorrington, Schomaker,
  Hoffmann, Keiser, Fox, Wood, Prozesky, Giddy, et~al.}{Johnson
  et~al.}{2013}]{johnson:2013:ARTe0}
Johnson, L.~F., J.~Mossong, R.~E. Dorrington, M.~Schomaker, C.~J. Hoffmann,
  O.~Keiser, M.~P. Fox, R.~Wood, H.~Prozesky, J.~Giddy, et~al. (2013).
\newblock Life expectancies of {S}outh {A}frican adults starting antiretroviral
  treatment: collaborative analysis of cohort studies.
\newblock {\em PLoS Medicine\/}~{\em 10\/}(4), e1001418.

\bibitem[\protect\citeauthoryear{Lee and Miller}{Lee and
  Miller}{2001}]{lee:2001:evaluatingLeeCarter}
Lee, R. and T.~Miller (2001).
\newblock Evaluating the performance of the {L}ee-{C}arter method for
  forecasting mortality.
\newblock {\em Demography\/}~{\em 38\/}(4), 537--549.

\bibitem[\protect\citeauthoryear{Lee and Carter}{Lee and
  Carter}{1992}]{leecarter:1992}
Lee, R.~D. and L.~R. Carter (1992).
\newblock Modeling and forecasting {US} mortality.
\newblock {\em Journal of the American Statistical Association\/}~{\em
  87\/}(419), 659--671.

\bibitem[\protect\citeauthoryear{Lutz, Sanderson, and Scherbov}{Lutz
  et~al.}{1998}]{lutz:1998:expert}
Lutz, W., W.~C. Sanderson, and S.~Scherbov (1998).
\newblock Expert-based probabilistic population projections.
\newblock {\em Population and Development Review\/}~{\em 24}, 139--155.

\bibitem[\protect\citeauthoryear{Mills, Bakanda, Birungi, Chan, Ford, Cooper,
  Nachega, Dybul, and Hogg}{Mills et~al.}{2011}]{mills:2011:ARTe0}
Mills, E.~J., C.~Bakanda, J.~Birungi, K.~Chan, N.~Ford, C.~L. Cooper, J.~B.
  Nachega, M.~Dybul, and R.~S. Hogg (2011).
\newblock Life expectancy of persons receiving combination antiretroviral
  therapy in low-income countries: a cohort analysis from {U}ganda.
\newblock {\em Annals of Internal Medicine\/}~{\em 155\/}(4), 209--216.

\bibitem[\protect\citeauthoryear{Raftery, Chunn, Gerland, and
  {\v{S}}ev{\v{c}}{\'\i}kov{\'a}}{Raftery et~al.}{2013}]{raftery:2013:e0}
Raftery, A.~E., J.~L. Chunn, P.~Gerland, and H.~{\v{S}}ev{\v{c}}{\'\i}kov{\'a}
  (2013).
\newblock Bayesian probabilistic projections of life expectancy for all
  countries.
\newblock {\em Demography\/}~{\em 50\/}(3), 777--801.

\bibitem[\protect\citeauthoryear{Raftery, Lalic, and Gerland}{Raftery
  et~al.}{2014}]{raftery:2014:jointe0}
Raftery, A.~E., N.~Lalic, and P.~Gerland (2014).
\newblock Joint probabilistic projection of female and male life expectancy.
\newblock {\em Demographic Research\/}~{\em 30}, 795.

\bibitem[\protect\citeauthoryear{Stanecki, Garnett, and Ghys}{Stanecki
  et~al.}{2012}]{stanecki:2012:HIVproj}
Stanecki, K., G.~P. Garnett, and P.~D. Ghys (2012).
\newblock Developments in the field of {HIV} estimates: methods, parameters and
  trends.
\newblock {\em Sexually Transmitted Infections\/}~{\em 88}, i1--i2.

\bibitem[\protect\citeauthoryear{{Statistics South Africa}}{{Statistics South
  Africa}}{2010}]{ssa:2010}
{Statistics South Africa} (2010).
\newblock Mid-year population estimates 2010.
\newblock http://www.statssa.gov.za/publications/P0302/P03022010.pdf.

\bibitem[\protect\citeauthoryear{{Statistics South Africa}}{{Statistics South
  Africa}}{2011}]{ssa:2011}
{Statistics South Africa} (2011).
\newblock Mid-year population estimates 2011.
\newblock http://www.statssa.gov.za/publications/P0302/P03022011.pdf.

\bibitem[\protect\citeauthoryear{{Statistics South Africa}}{{Statistics South
  Africa}}{2013}]{ssa:2013}
{Statistics South Africa} (2013).
\newblock Mid-year population estimates 2013.
\newblock http://www.statssa.gov.za/publications/P0302/P03022013.pdf.

\bibitem[\protect\citeauthoryear{{Statistics South Africa}}{{Statistics South
  Africa}}{2014}]{ssa:2014}
{Statistics South Africa} (2014).
\newblock Mid-year population estimates 2014.
\newblock https://www.statssa.gov.za/publications/P0302/P03022014.pdf.

\bibitem[\protect\citeauthoryear{{Statistics South Africa}}{{Statistics South
  Africa}}{2015}]{ssa:2015}
{Statistics South Africa} (2015).
\newblock Mid-year population estimates 2015.
\newblock https://www.statssa.gov.za/publications/P0302/P03022015.pdf.

\bibitem[\protect\citeauthoryear{Stover, Brown, and Marston}{Stover
  et~al.}{2012}]{stover:2012:spectrumEPP}
Stover, J., T.~Brown, and M.~Marston (2012).
\newblock Updates to the {Spectrum/Estimation and Projection Package (EPP)}
  model to estimate {HIV} trends for adults and children.
\newblock {\em Sexually Transmitted Infections\/}~{\em 88\/}(Suppl 2),
  i11--i16.

\bibitem[\protect\citeauthoryear{{UNAIDS}}{{UNAIDS}}{2014}]{spectrum:2014:guide}
{UNAIDS} (2014, updated Janurary).
\newblock {\em Quick Start Guide for {Spectrum} (Downloaded December 2015)}.
\newblock {UNAIDS}.

\bibitem[\protect\citeauthoryear{{United Nations}}{{United
  Nations}}{2013}]{un:WPP2012}
{United Nations} (2013).
\newblock {\em World Population Prospects: The 2012 Revision}.
\newblock United Nations, Department of Economic and Social Affairs, Population
  Division New York, NY, USA.

\bibitem[\protect\citeauthoryear{{United Nations}}{{United
  Nations}}{2015}]{un:WPP2015}
{United Nations} (2015).
\newblock {\em World Population Prospects: The 2015 Revision}.
\newblock United Nations, Department of Economic and Social Affairs, Population
  Division New York, NY, USA.

\end{thebibliography}

\clearpage
\section*{Appendix A: Female Life Expectancy and Adult HIV Prevalence Projections}
In the left panel of every row below are projections of life expectancy using our model \eqref{eq:e0HnA} for all countries we treat as generalized epidemic countries. In the right panel, our observed trajectories of HIV prevalence and our projections of HIV prevalence made using a single trajectory of HIV prevalence for each country allowing the EPP projections to give uncertainty. The format of these figures is the same as that of Figures 4-7 in Section~\ref{sec:CaseStudies}. 
\includepdf[pages=-]{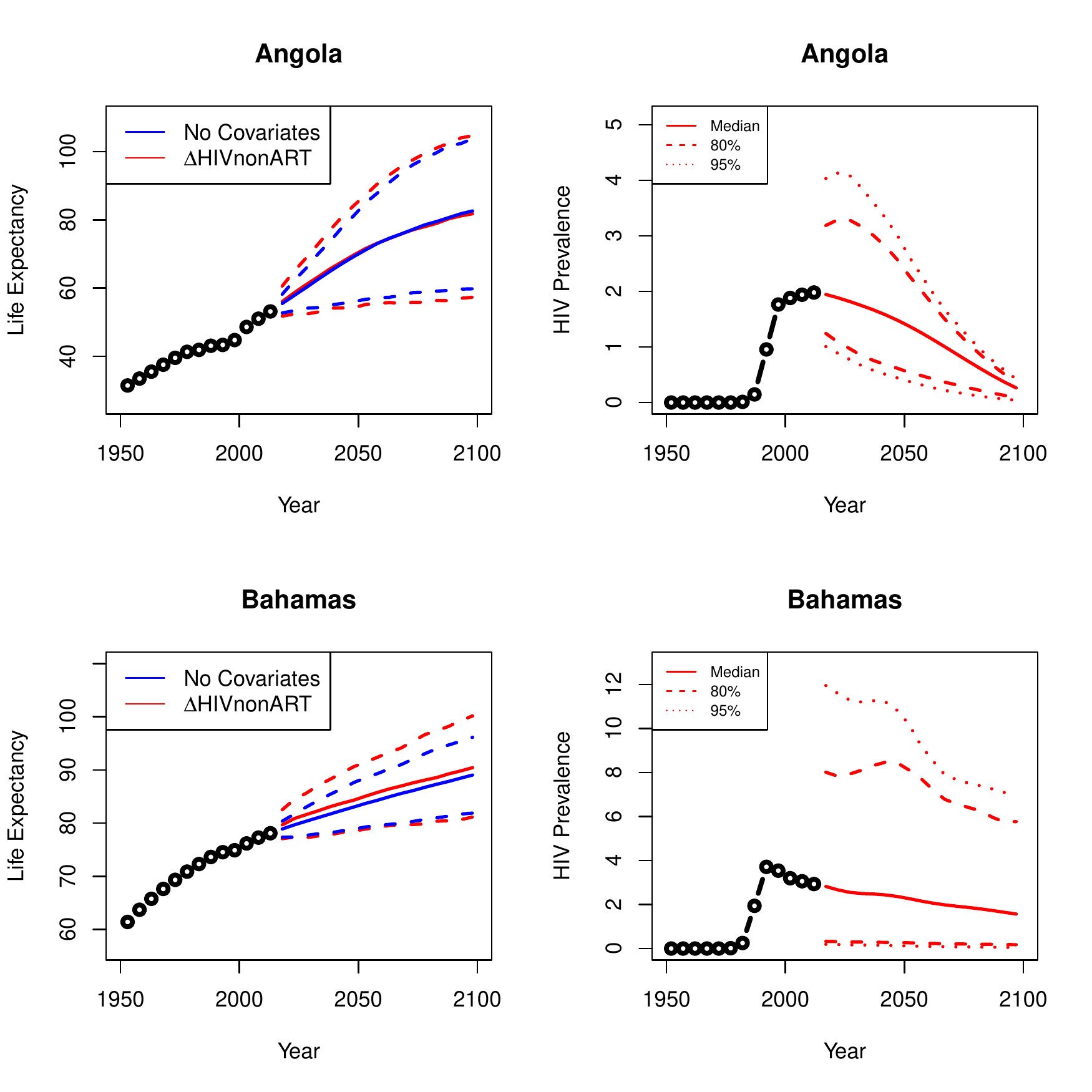}

\section*{Appendix B: Male Life Expectancy and Adult HIV Prevalence Projections}

\begin{table}
\centering
\caption{Out of Sample Validation Results for male life expectancy projected using the gap model in \citep{raftery:2014:jointe0}. The first column represents the set of countries used in the subsequent calibration calculations. The second column represents the time period of data used to fit the model, and the third column represents the time periods used in validation. The fourth column represents the number of countries used in validation. In the fifth column, ``No Covariates" represents the model in \eqref{eq:e0noHIV} and ``$\Delta HnA$" represents the model in \eqref{eq:e0HnA}. The sixth column contains the MAE as defined in Section~\ref{subsec:Validation}. The seventh and eighth columns contain coverage metrics for the 80\% and 95\% predictive intervals respectively.}
\vspace{0.2in}
\begin{tabular}{|c|c|c|c|c|c|cc|}
\hline
\multicolumn{1}{|c|}{Countries} & \multicolumn{1}{c|}{Training} &\multicolumn{1}{c|}{Test} &\multicolumn{1}{c|}{$n$} &\multicolumn{1}{c|}{Model} & \multicolumn{1}{c|}{MAE} & \multicolumn{2}{c|}{Coverage}  \\ \cline{7-8}
 & Period & Period & & & & 80\% & 95\%   \\
\hline
\multirow{8}{*}{HIV} & \multirow{2}{*}{1950-2005} & \multirow{2}{*}{2005-2015} & \multirow{2}{*}{69} & No Covariates & 3.74 & 0.36 & 0.51 \\
& & & & $\Delta HnA$ & 2.49 & 0.75 & 0.84 \\ \cline{2-8}
& \multirow{2}{*}{1950-2005} & \multirow{2}{*}{2005-2010} & \multirow{2}{*}{40} & No Covariates & 2.69 & 0.40 & 0.55  \\
& & & & $\Delta HnA$ & 2.41 & 0.73 & 0.80  \\ \cline{2-8}
& \multirow{2}{*}{1950-2005} & \multirow{2}{*}{2010-2015} & \multirow{2}{*}{29} & No Covariates & 5.19 & 0.31 & 0.45  \\
& & & & $\Delta HnA$  & 2.60 & 0.79 & 0.90 \\ \cline{2-8}
& \multirow{2}{*}{1950-2010} & \multirow{2}{*}{2010-2015} & \multirow{2}{*}{29} & No Covariates & 3.22 & 0.38 & 0.55 \\
& & &  & $\Delta HnA$  & 1.81 & 0.76 & 0.86  \\ \cline{2-8}
\hline
\multirow{8}{*}{All} & \multirow{2}{*}{1950-2005} & \multirow{2}{*}{2005-2015} & \multirow{2}{*}{390} & No Covariates & 1.46 & 0.79 & 0.88  \\
& & & & $\Delta HnA$ & 1.23 & 0.86 & 0.94 \\ \cline{2-8}
& \multirow{2}{*}{1950-2005} & \multirow{2}{*}{2005-2010} & \multirow{2}{*}{201} & No Covariates & 1.11 & 0.80 & 0.89  \\
& & & & $\Delta HnA$ & 1.05 & 0.87 & 0.94 \\ \cline{2-8}
& \multirow{2}{*}{1950-2005} & \multirow{2}{*}{2010-2015} & \multirow{2}{*}{189} & No Covariates & 1.83 & 0.77 & 0.88  \\
& & & & $\Delta HnA$ & 1.43 & 0.86 & 0.95 \\ \cline{2-8}
& \multirow{2}{*}{1950-2010} & \multirow{2}{*}{2010-2015} & \multirow{2}{*}{189} & No Covariates & 1.20 & 0.76 & 0.86 \\
& & & & $\Delta HnA$ & 0.99 & 0.82 & 0.90 \\ \cline{2-8}
\hline
\end{tabular}
\label{tab:modelvalidM}
\end{table}
\clearpage

\includepdf[pages=-]{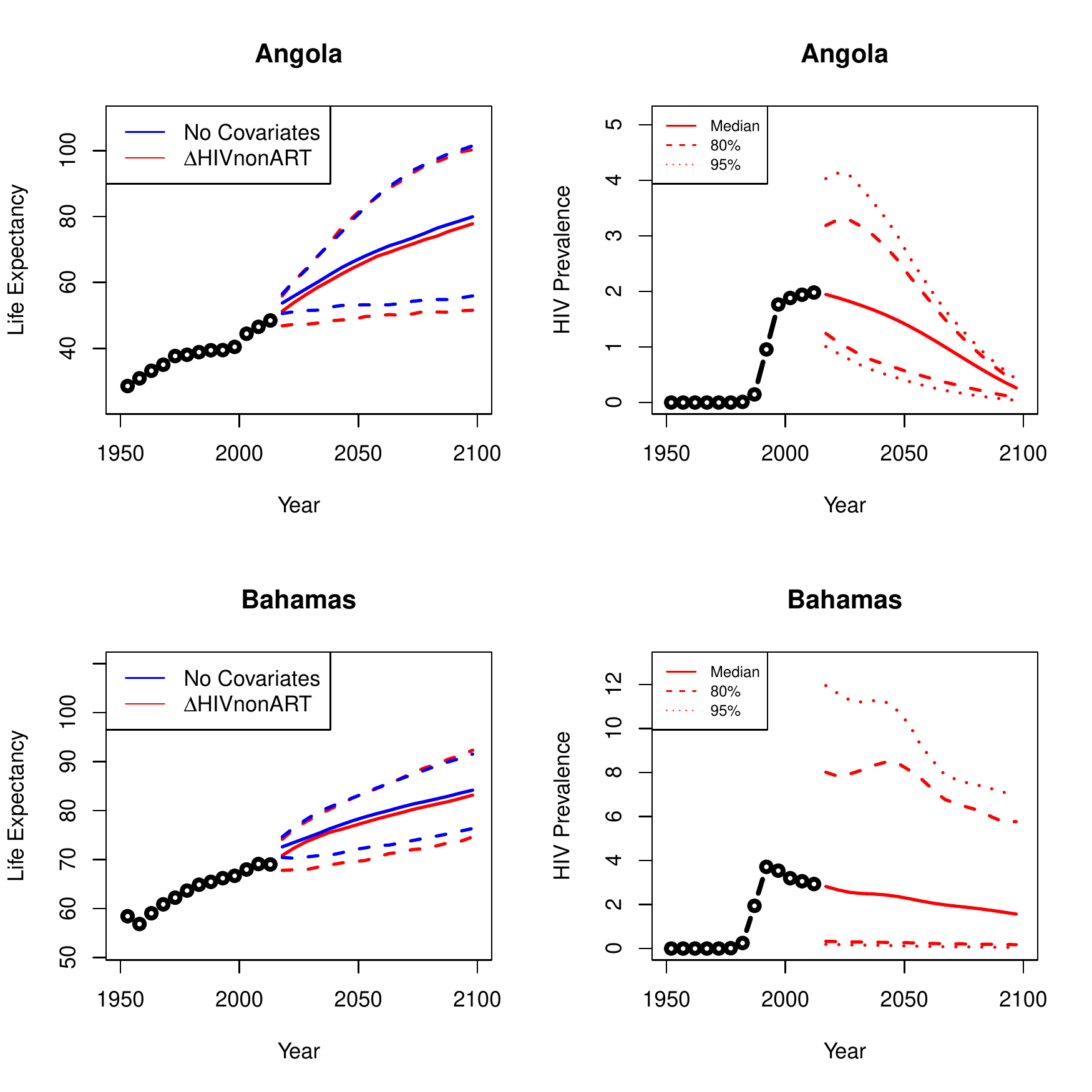}

\end{document}